\documentclass[preprint,12pt]{article}
\usepackage{amssymb}
\usepackage{amsxtra}
\usepackage{amsmath}
\usepackage{amstext}
\usepackage{amsthm}
\usepackage{amsbsy}
\usepackage{latexsym}
\usepackage{amscd}
\usepackage{eucal}
\usepackage[dvips]{graphicx}
\usepackage{graphics}
\usepackage{float}
\usepackage{anysize}
\usepackage{xcolor}
\usepackage[font=small,labelfont=bf]{caption}

\usepackage{authblk}
\usepackage{amsfonts}
\usepackage[english]{babel}

\title{Higher-order statistical correlations and mutual information among particles in a quantum well}
\author[a]{V.S. Y{\'e}pez}
\author[a]{R.P. Sagar}
\author[a,b]{H.G. Laguna\footnote{Corresponding author: hlag@ciencias.unam.mx}}
\affil[a]{Departamento de Qu{\'i}mica, Universidad Aut{\'o}noma Metropolitana, San Rafael Atlixco No. 186, Iztapalapa, 09340, Ciudad de M{\'e}xico, M{\'e}xico.}
\affil[b]{Departamento de Matem{\'a}ticas, Facultad de Ciencias; y Centro de Ciencias de la Complejidad, Universidad Nacional Aut{\'o}noma de M{\'e}xico, Circuito Exterior, Ciudad Universitaria, Ciudad de M{\'e}xico, 04510, M{\'e}xico.}
\date{\today}

\begin{document}
\maketitle

\begin{abstract}
The influence of wave function symmetry on statistical correlation is studied for the case of three non-interacting spin-free quantum particles in a unidimensional box, in position and in momentum space.
Higher-order statistical correlations occurring among the three particles in this quantum system is quantified via higher-order mutual information and compared to the correlation between pairs of variables in this model, and to the correlation in the two-particle system. The results for the higher-order mutual information show that there are states where the symmetric wave functions are more correlated than the antisymmetric ones with same quantum numbers. This holds in position as well as in momentum space. This behavior is opposite to that observed for the correlation between pairs of variables in this model, and the two-particle system, where the antisymmetric wave functions are in general more correlated. These results are also consistent with those observed in a system of three uncoupled oscillators. The use of higher-order mutual information as a correlation measure, is monitored and examined by considering a superposition of states or systems with two Slater determinants.
\end{abstract}


\section{Introduction}

The study of the effects of indistinguishability on the statistical correlation between particles in quantum systems emanates from the need to understand the characteristics that the wave functions present in such systems \cite{tichy2012,boquillon2013,bose2013,ma2014,gong2014,tichy2014,lopes2015}. An understanding of how wave function properties influence the correlation among quantum particles is important for understanding quantum phenomena, especially important when addressing the present and future miniaturization of technological devices.
 
The symmetrization of wave functions arising from the indistinguishability requirement they must fulfill is a source of correlation between positions or momenta of the particles involved since the wave function cannot be written in a separable form. This type of correlation is distinct from the correlation that arises from a superposition of states as we shall examine in the last section. 

The concept of examining correlation via the statistical correlation, in electronic quantum systems, was introduced into the literature \cite{wigner,wigner2} and has been examined via the correlation coefficient \cite{kutzelnigg}. Our interests lie in determining the role of statistical correlation due to wave function symmetry in the distribution functions in position and in momentum space. In this work, we study the statistical correlation in the three-particle in a box system, to examine differences between symmetric and antisymmetric wave functions which for these spinless systems would be analogs of bosonic and fermionic wave functions. We first examine the correlation between pairs in the reduced distributions. Secondly, and most importantly, we then analyze the correlation among the three variables or particles by examining the higher-order mutual information and related measures. 

Symmetric and antisymmetric wave functions for this three-particle spinless non-interacting system can be built from orbitals which are solutions to the one-particle in a box system of length $L$.

In position space, these orbitals are 

\begin{equation}
\psi_n(x)=\sqrt{\frac{2}{L}}  \mathit{sin} \bigg(\frac{n\pi x}{L}\bigg) \qquad \qquad 0\leq x \leq L.
\label{orbitalx}
\end{equation}
The corresponding orbitals in momentum space \cite{belloni} are obtained using the Dirac-Fourier transform (with $\hbar =1$),
\begin{eqnarray}
\nonumber
\phi_n(p) = & \sqrt{\frac{1}{2\pi}} \int_0^L \psi_n(x)\;e^{-ipx}\;dx & -\infty\leq p\leq \infty \\
= & (- i) \sqrt{ \frac{L}{\pi }} e^{-ip L/2} {\bigg [} e^{in\pi /2} \frac{\sin{[(pL  - n\pi)/2]}}{(pL - n\pi)} - e^{-in\pi /2} \frac{\sin{[(pL + n\pi)/2]}}{(pL + n\pi)} {\bigg ]}. &
\label{orbitalp}
\end{eqnarray}

Wave functions in position or in momentum space can be built from these orbitals using permanents (symmetric) or Slater determinants (antisymmetric) as

\begin{eqnarray}
\nonumber
\Psi_{n_1,n_2,n_3}(x_1,x_2,x_3)= \frac{1}{\sqrt{6}} \ \ |\psi_{n_1}(x_1) \; \psi_{n_2}(x_2) \; \psi_{n_3}(x_3)| \\
\Phi_{n_1,n_2,n_3}(p_1,p_2,p_3)= \frac{1}{\sqrt{6}} \ \ |\phi_{n_1}(p_1) \; \phi_{n_2}(p_2) \; \phi_{n_3}(p_3)|.
\label{wfns}
\end{eqnarray}
For symmetric functions or permanents with the three particles in the same one-particle state one has to divide by $\sqrt{6}$ while for two particles in the same one-particle state one divides by $\sqrt{2}$ to correctly normalize the wave function. We will not consider such symmetric functions since their antisymmetric counterparts are zero. Furthermore, three particles in the same one-particle state would yield a statistical correlation of zero.

The particle in a box (PIAB) is a simple model but has wide application. For example, in the modeling of electrons in conjugated molecules \cite{hobey}, the behavior of electrons in metals and serves as a model in the design of devices such as quantum wells.

Information theory applied to the study of quantum systems offers the possibility of gaining insights into the behavior of quantum systems. This theory provides measures of uncertainty and correlation known as the Shannon entropy and mutual information. The Shannon entropy \cite{shannon} has been applied to study the uncertainty in the particle in a box model in position and in momentum space \cite{majernik1,majernik1a,majernik2,sanchezruiz}. The study of Shannon entropies in quantum systems is of current interest \cite{olendski2016, olendski2015, mukerjee2016, mukerjee2015, ghafourian2016, najafizade2016, fotue2015, sun2013, dong2014}.

\section{Shannon entropies}

The Shannon information entropy is defined in position or in momentum space as
\begin{equation}
s_x= - \int \rho (x) \ln \rho (x) \; dx \quad \quad s_p= -\int \pi (p) \ln \pi (p) \; dp,
\label{shanentropy}
\end{equation}
where $\rho (x)$ and $\pi (p)$ are the respective position and momentum space one-particle (variable) densities normalized to unity. In one-particle systems, these
densities are the squared moduli of the wave functions while for $N$-particle systems they are the reduced densities obtained by integration over $N-1$ variables. All
integrals in position space are over the $[0,L]$ interval, while in momentum space they are over $[-\infty,\infty]$, as seen in Eqs. (\ref{orbitalx}) and (\ref{orbitalp}).

The interest in the Shannon entropies of both position and momentum space stems from the existence of the entropic uncertainty relationship \cite{bbm}
\begin{equation}
s_x + s_p \ge 1 + \ln \pi.
\label{bbm}
\end{equation}
The entropic uncertainty relationship has been shown to be superior to the one based on standard deviations \cite{deutsch,uffink,wehner,zhang}. The entropy sum, $s_x+s_p$, can be interpreted as the Shannon entropy of a separable phase-space distribution, $\rho (x)\pi (p)$.

Two-particle or pair Shannon entropies can also be considered in $N$-particle systems. These are defined as
\begin{eqnarray}
\nonumber
s_{\Gamma}= - \int \Gamma (x_1,x_2) \ln \Gamma (x_1,x_2) \; dx_1dx_2 \\
 s_{\Pi}= - \int \Pi (p_1,p_2) \ln \Pi (p_1,p_2) \; dp_1dp_2 
\label{twoshan}
\end{eqnarray}
where $\Gamma(x_1,x_2)$ and $\Pi(p_1,p_2)$ are the two-particle densities in position and in momentum space, normalized to unity. In two-particle systems, these
are the squared moduli of the wave functions while for three or more particle systems, they are reduced densities.

In the case of three-particle systems, such as the ones of interest here, three-particle Shannon entropies can be defined in terms of the wave functions as
\begin{eqnarray}
\nonumber
s_{\Psi}= -\int & |\Psi_{n_1,n_2,n_3}(x_1,x_2,x_3)|^2 \ln |\Psi_{n_1,n_2,n_3}(x_1,x_2,x_3)|^2
\; \; dx_1dx_2dx_3 \\
s_{\Phi}= -\int &|\Phi_{n_1,n_2,n_3}(p_1,p_2,p_3)|^2 \ln |\Phi_{n_1,n_2,n_3}(p_1,p_2,p_3)|^2 \; \; dp_1dp_2dp_3.
\label{threeshan}
\end{eqnarray}
In systems with more than three particles, the squared moduli of the wave functions would be replaced with reduced densities in the above expressions.

The Shannon entropies at each level are measures of the uncertainties in the underlying distributions. These measures increase as the densities delocalize
and decrease as the densities localize or the uncertainty diminishes. There are other measures that have been employed to measure uncertainty. Our interest
in the Shannon entropies lie in their connection to measures of statistical correlation as we shall see in the next section. We emphasize that these uncertainty measures do not quantify uncertainty over states but rather the spatial uncertainty in position or in momentum space.

\section{Mutual Information}

The interest in the one- and two-particle (variable) Shannon entropies is that they form the basis of the definition of mutual information \cite{cover}.
In position and in momentum space these are \cite{sagar_jcp_2005,sagar_jcp_2006}
\begin{eqnarray}
\nonumber
I_x= \int \Gamma (x_1,x_2) \ln \Bigg [ \frac{\Gamma (x_1,x_2)}{\rho (x_1)\rho (x_2)} \Bigg ] \; dx_1dx_2 = 2s_x-s_{\Gamma} \ge 0 \\
\nonumber
I_p= \int \Pi (p_1,p_2) \ln \Bigg [ \frac{\Pi (p_1,p_2)}{\pi (p_1)\pi (p_2)} \Bigg ] \; dp_1dp_2 = 2s_p-s_{\Pi} \ge 0 .\\
\label{mutinf}
\end{eqnarray}

For the three-particle systems under consideration here, $\Gamma (x_1,x_2)$ and $\Pi (p_1,p_2)$ are the two-particle reduced densities 
\begin{eqnarray}
\nonumber
\Gamma (x_1,x_2) = \Gamma_{n_1,n_2,n_3} (x_1,x_2) = \int |\Psi_{n_1,n_2,n_3}(x_1,x_2,x_3)|^2 \; dx_3 ,\\ 
\nonumber
\Pi (p_1,p_2) = \Pi_{n_1,n_2,n_3} (p_1,p_2) = \int |\Phi_{n_1,n_2,n_3}(p_1,p_2,p_3)|^2 \; dp_3. \\
\label{twodensity}
\end{eqnarray}
Note that since the particles are indistinguishable, the integration can be performed over any one of the three variables. Furthermore, these reduced densities depend on all three quantum numbers, as shown above.

The one-particle (variable) reduced densities are similarly defined as
\begin{eqnarray}
\nonumber
\rho (x) = \rho_{n_1,n_2,n_3}(x) = \int \Gamma (x,x_2) \; dx_2, \\
 \pi (p)= \pi_{n_1,n_2,n_3}(p)= \int \Pi (p,p_2) \; dp_2 ,
\label{onedensity}
\end{eqnarray}
where we have dropped the superscript on variable one to emphasize the indistinguishability. Again these reduced densities depend on $n_{1}$, $n_{2}$ and $n_{3}$.

Mutual information is a measure of the statistical correlation or dependency between pairs of variables. It is zero for independent variables or
distributions that are separable and increases as the dependency between variables increases. It is a more general measure of correlation than the
correlation coefficient which is only able to detect linear dependencies. This has been observed in applications to quantum systems \cite{laguna_pra_2011}.

The effects of wave function sysmmetry on the statistical correlation between particles has been discussed in the case of two particles in a box \cite{laguna_jpa_2011}. Systems where the particles interact via harmonic potentials have also been examined \cite{laguna_physica_2013,entropy}. The results obtained showed that the antisymmetric wave functions were in general more correlated than the symmetric ones \cite{laguna_jpa_2011} with the same quantum numbers. Furthermore, the mutual information between pairs of variables was shown to be capable of detecting correlation in cases where the correlation coefficient yielded values of zero for the linear correlation. One goal of the present work is to determine how the correlation patterns change with the addition of more particles to the system.

\section{Mutual Information with three variables}

In systems with three particles or variables, there are several definitions that can be used where the correlation in the system is partitioned differently.
First, one can define in each space
\begin{equation}
I_{3x}= \int |\Psi_{n_1,n_2,n_3}(x_1,x_2,x_3)|^2 \ln \Bigg [ \frac{|\Psi_{n_1,n_2,n_3}(x_1,x_2,x_3)|^2}{\rho (x_1)\rho (x_2)\rho (x_3)} \Bigg ] \: dx_1dx_2dx_3 = 3s_x-s_{\Psi}
\label{mi1x}
\end{equation}
\begin{equation}
I_{3p}= \int |\Phi_{n_1,n_2,n_3}(p_1,p_2,p_3)|^2 \ln \Bigg [ \frac{|\Phi_{n_1,n_2,n_3}(p_1,p_2,p_3)|^2}{\pi (p_1)\pi (p_2)\pi (p_3)} \Bigg ] \: dp_1dp_2dp_3 =3s_p-s_{\Phi}.
\label{mi1p}
\end{equation}
These quantities measure the total correlations in the three variable system and include all correlations between pairs as well as
the correlation shared among all three variables.

Another definition that can be used is \cite{entropy}
\begin{equation}
I_{\rho,\Gamma}= \int |\Psi_{n_1,n_2,n_3}(x_1,x_2,x_3)|^2 \ln \Bigg [ \frac{|\Psi_{n_1,n_2,n_3}(x_1,x_2,x_3)|^2}{\rho (x_1)\Gamma (x_2,x_3)} \Bigg ] \; dx_1dx_2dx_3 =s_x+s_{\Gamma}-s_{\Psi}
\label{mi2x}
\end{equation}
\begin{equation}
I_{\pi, \Pi}= \int |\Phi_{n_1,n_2,n_3}(p_1,p_2,p_3)|^2 \ln \Bigg [ \frac{|\Phi_{n_1,n_2,n_3}(p_1,p_2,p_3)|^2}{\pi (p_1)\Pi (p_1,p_2)} \Bigg ]  \; dp_1dp_2dp_3 =s_p+s_{\Pi}-s_{\Phi}.
\label{mi2p}
\end{equation}
This quantity can be interpreted as measuring the correlation between one variable (particle) and a pair of variables.

Likewise,
\begin{equation}
I_{\Gamma,\Gamma}= \int |\Psi_{n_1,n_2,n_3}(x_1,x_2,x_3)|^2 \ln \Bigg [ \frac{|\Psi_{n_1,n_2,n_3}(x_1,x_2,x_3)|^2 \rho(x_1)}{\Gamma (x_1,x_2)\Gamma (x_1,x_3)} \Bigg ] \; dx_1dx_2dx_3 =2s_{\Gamma}-s_x-s_{\Psi}
\label{mi3x}
\end{equation}
\begin{equation}
I_{\Pi, \Pi}= \int |\Phi_{n_1,n_2,n_3}(p_1,p_2,p_3)|^2 \ln \Bigg [ \frac{|\Phi_{n_1,n_2,n_3}(p_1,p_2,p_3)|^2 \pi (p_1)}{\Pi (p_1,p_2)\Pi (p_1,p_3)} \Bigg ] \; dp_1dp_2dp_3 =2s_{\Pi}-s_p-s_{\Phi}.
\label{mi3p}
\end{equation}
measures the correlation between pairs of variables.

\subsection{Higher-order mutual information}

Perhaps of greatest interest is a measure which is capable of detecting higher-order correlation or the dependency among the
three variables as a group. One can define a higher-order mutual information \cite{cover,matsu,matsub,killian,somani} as
\begin{eqnarray}
\nonumber
I_x^3 & =& \int dx_1dx_2dx_3 \; \; |\Psi_{n_1,n_2,n_3}(x_1,x_2,x_3)|^2
 \; \ln \Bigg [ \frac{|\Psi_{n_1,n_2,n_3}(x_1,x_2,x_3)|^2 \rho(x_1) \rho (x_2) \rho (x_3)}{\Gamma (x_1,x_2)\Gamma (x_1,x_3)\Gamma (x_2,x_3)} \Bigg ]  \\
& =& 3s_{\Gamma}-3s_x-s_{\Psi}
\label{mi4x}
\end{eqnarray}
\begin{eqnarray}
\nonumber
I_p^3 &=& \int dp_1dp_2dp_3 \; \; |\Phi_{n_1,n_2,n_3}(p_1,p_2,p_3)|^2 \; \ln \Bigg [ \frac{|\Phi_{n_1,n_2,n_3}(p_1,p_2,p_3)|^2 \pi (p_1) \pi (p_2) \pi (p_3)}{\Pi (p_1,p_2)\Pi (p_1,p_3)\Pi (p_2,p_3)} \Bigg ] \\
& =& 3s_{\Pi}-3s_p-s_{\Phi}.
\label{mi4p}
\end{eqnarray}
These expressions do not include a minus sign in the definition, as used elsewhere \cite{matsu,matsub}, since it is closer to the definition of the pair mutual information.

Note that the second equality in the above expressions, and also those of the other previous mutual information measures, are a consequence of the indistinguishability requirements of quantum systems.
For example, Eq. (\ref{mi4x}) would be 
\begin{equation}
I_x^3=s_{\Gamma_{x_1,x_2}}+ s_{\Gamma_{x_1,x_3}}+s_{\Gamma_{x_2,x_3}}- \Big [s_{x_1}+s_{x_2}+s_{x_3}+s_{\Psi} \Big ]
\label{disting}
\end{equation}
for a distinguishable system.

\begin{figure}
\includegraphics[width=\columnwidth]{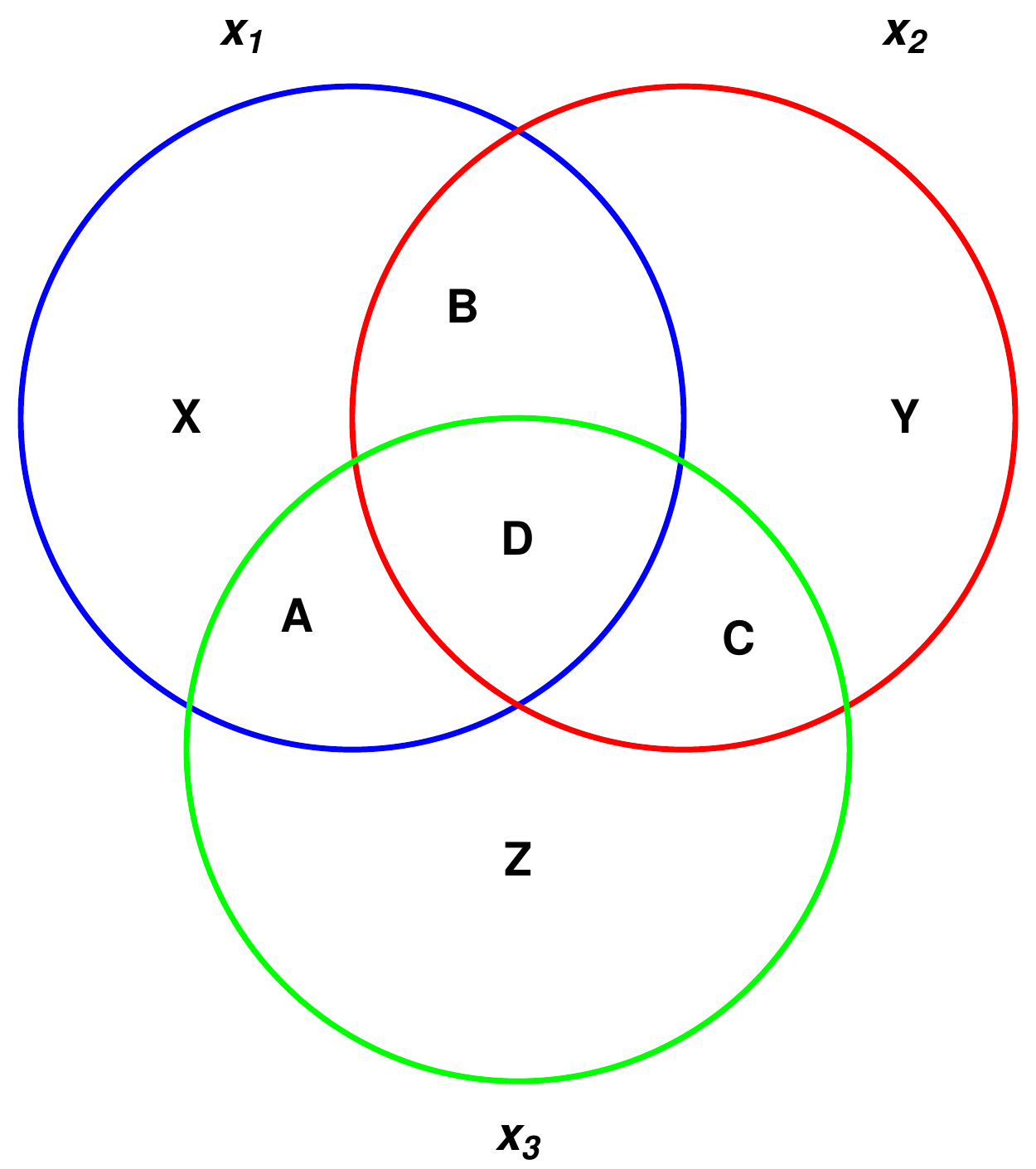}
\caption{\label{diagram} \small{Diagram showing a schematic representation of the intersections between and among the three variables: blue circle ($x_1$), red circle ($x_2$), green circle ($x_3$). All regions are delineated by the three different colors.}}
\end{figure}

Eq. (\ref{disting}) can be understood from inspection of Fig. (\ref{diagram}).
Regions {\bf A} plus {\bf D}, the region shared by $x_1$ and $x_3$, represents the overlap or intersection between $x_1$ and $x_3$; {\bf B} plus {\bf D},
the intersection between $x_1$ and $x_2$ while {\bf D} plus {\bf C}, the intersection between $x_2$ and $x_3$. These intersections, or correlations, are measured
by the pair mutual information. The intersection among $all$ three variables is represented by region {\bf D}, the region common to all three circles, and is measured by the higher-order mutual information.

In entropic terms, the sum of all regions is $S_{\Psi}$. $S_{\Gamma{x_1,x_2}}$ is the sum of all regions belonging to the blue and red circles, $s_{\Gamma{x_1,x_3}}$, the regions belonging
to the blue and green circles, and $S_{\Gamma{x_2,x_3}}$, the regions of the red and green circles. $s_{x_1}$, $s_{x_2}$ and $s_{x_3}$ are the areas belonging to the blue, red and green circles respectively. One can see that summing and subtracting the different regions corresponding to the different terms in Eq. (\ref{disting}) will yield the (minus) {\bf D} region. Likewise, $I_{3x}$ corresponds to the {\bf BADDC} regions ({\bf D} is counted twice). $I_{\rho, \Gamma}$ consists of the {\bf BAD} regions while $I_{\Gamma, \Gamma}$ is composed of the {\bf C} region.

One can also establish a hierarchy among the mutual information measures with three variables and the higher-order mutual information. The differences between
Eqs. (\ref{mi1x}) and (\ref{mi2x}), Eqs. (\ref{mi2x}) and (\ref{mi3x}) and Eqs. (\ref{mi3x}) and (\ref{mi4x}), are all equal to $I_x$, which is greater than or equal to zero.
Thus
\begin{equation}
I_{3x} \ge I_{\rho, \Gamma} \ge I_{\Gamma, \Gamma} \ge I_x^3.
\label{hierx}
\end{equation}
In a similar fashion in momentum space,
\begin{equation}
I_{3p} \ge I_{\pi, \Pi} \ge I_{\Pi, \Pi} \ge I_p^3.
\label{hierp}
\end{equation}

The higher-order mutual information goes beyond what is captured between pairs and measures the correlation or dependency among the three variables. On the other hand, it can be negative \cite{cover}, unlike the mutual information between pairs of variables.
Much less work has been done on higher-order mutual information as compared to the pair mutual information which has seen wide application in a variety of scientific disciplines.
Also lacking is a full understanding and interpretation of the sign of higher-order mutual information.
Mutual information measures are given in natural units or nats since the units of the densities in the numerator and denominator of the logarithmic argument cancel. 

One could also consider measuring correlation by extending the concept of the correlation coefficient to a $ 3 \times 3$ covariance or correlation matrix for these three-variable systems. In general, these are symmetric matrices with elements $\sigma _{ij}=\sigma_{ji}$. In these systems with particle indistinguishability, not only would the matrices be symmetric, but each element would also be equal to all others. We did not pursue this avenue since it is not transparent how higher-order correlations could be obtained from these matrices.

Furthermore, the third-order cumulants, defined as $C_x = \langle x_1 x_2 x_3 \rangle - 3\langle x_i x_j \rangle \langle x \rangle + 2 \langle x \rangle ^{3}$ and $C_p = \langle p_1 p_2 p_3 \rangle - 3\langle p_i p_j \rangle \langle p \rangle + 2 \langle p \rangle ^{3}$, for indistinguishable particles, yield a value of zero in all cases. Thus the third-order cumulant is unable to quantify the higher-order correlation.

These statistical correlation measures are different from entanglement measures, based on the eigenvalues of the one-particle density matrix, which are representation free. The entanglement measures detect the correlations arising from interactions when one goes beyond the single Slater determinant approximation. Thus, for a single Slater determinant, there is no entanglement or correlation as understood in the context of electronic systems and the Hartree-Fock approximation \cite{huang2005,kais2007}. However, there is statistical correlation between particles in a single determinantal function, which is the object of this study. The information measures presented here can also be used to study interaction effects which are the result of physical potentials.

\section{Results and Discussion}

Information measures were calculated from the wave functions in Eq. (\ref{wfns}) by varying each of the quantum numbers $n_1, n_2,n_3=1-10$. The box length was fixed at $L=1$. The entropies, which form
the basis of the information measures, were calculated by one, two and three dimensional integrations using the Mathematica \cite{mathematica} package. The numerical values 
were checked against those generated from quadrature routines in FORTRAN. 

\subsection{Probability Densities}

The expressions for the two-particle (variable) reduced densities are
\begin{eqnarray}
\nonumber
\Gamma (x_1,x_2) = & \frac{1}{6} \Bigg [ \sum\limits_{i,j=1,i \neq j}^3 |\psi_{n_i}(x_1)|^2 |\psi_{n_j}(x_2)|^2 \pm \sum\limits_{i,j=1,i \neq j}^3 \psi_{n_i}^*(x_1)\psi_{n_j}^*(x_2)\psi_{n_j}(x_1)\psi_{n_i}(x_2) \Bigg ] \\
\Pi (p_1,p_2) = & \frac{1}{6} \Bigg [ \sum\limits_{i,j=1,i \neq j}^3 |\phi_{n_i}(p_1)|^2 |\phi_{n_j}(p_2)|^2 \pm \sum\limits_{i,j=1,i \neq j}^3 \phi_{n_i}^*(p_1)\phi_{n_j}^*(p_2)\phi_{n_j}(p_1)\phi_{n_i}(p_2) \Bigg ]
\label{twodens}
\end{eqnarray}
where the first summation includes the Hartree-like terms while the second the exchange terms. The $\pm$ corresponds to symmetric and antisymmetric wave functions respectively. These functions were chosen for simplicity and are not as general as in the case of Ref. \cite{fernandez2014}. We emphasize that our goal is to analyze differences in the information measures for the same chosen state.

\begin{figure}
\includegraphics[width=\columnwidth]{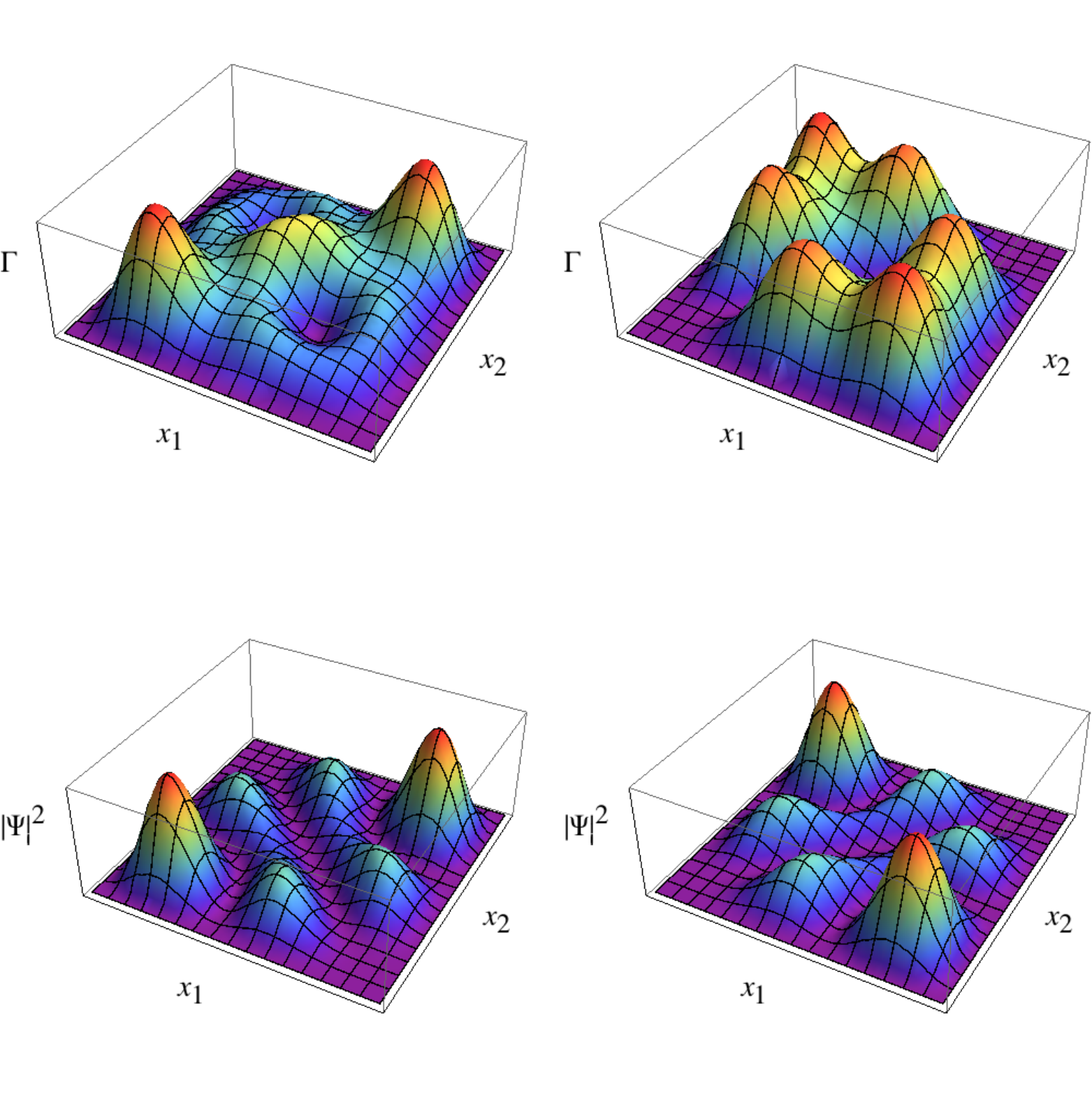}
\caption{\label{fig:distribucion HV} \small{Two-particle reduced densities in position space, $\Gamma (x_1,x_2)$ (top), with $(n_1, n_2, n_3)$= $(1,2,3)$ and  $L=1$ for symmetric (left) and antisymmetric (right) wave functions. The densities of the two-particle system,  $|\Psi (x_1,x_2)|^2$ (bottom), with $(n_1, n_2)$=$(2,3)$ and $L=1$ for symmetric (left) and antisymmetric (right) wave functions, are also shown.}}
\end{figure}

\begin{figure}
\includegraphics[width=\columnwidth]{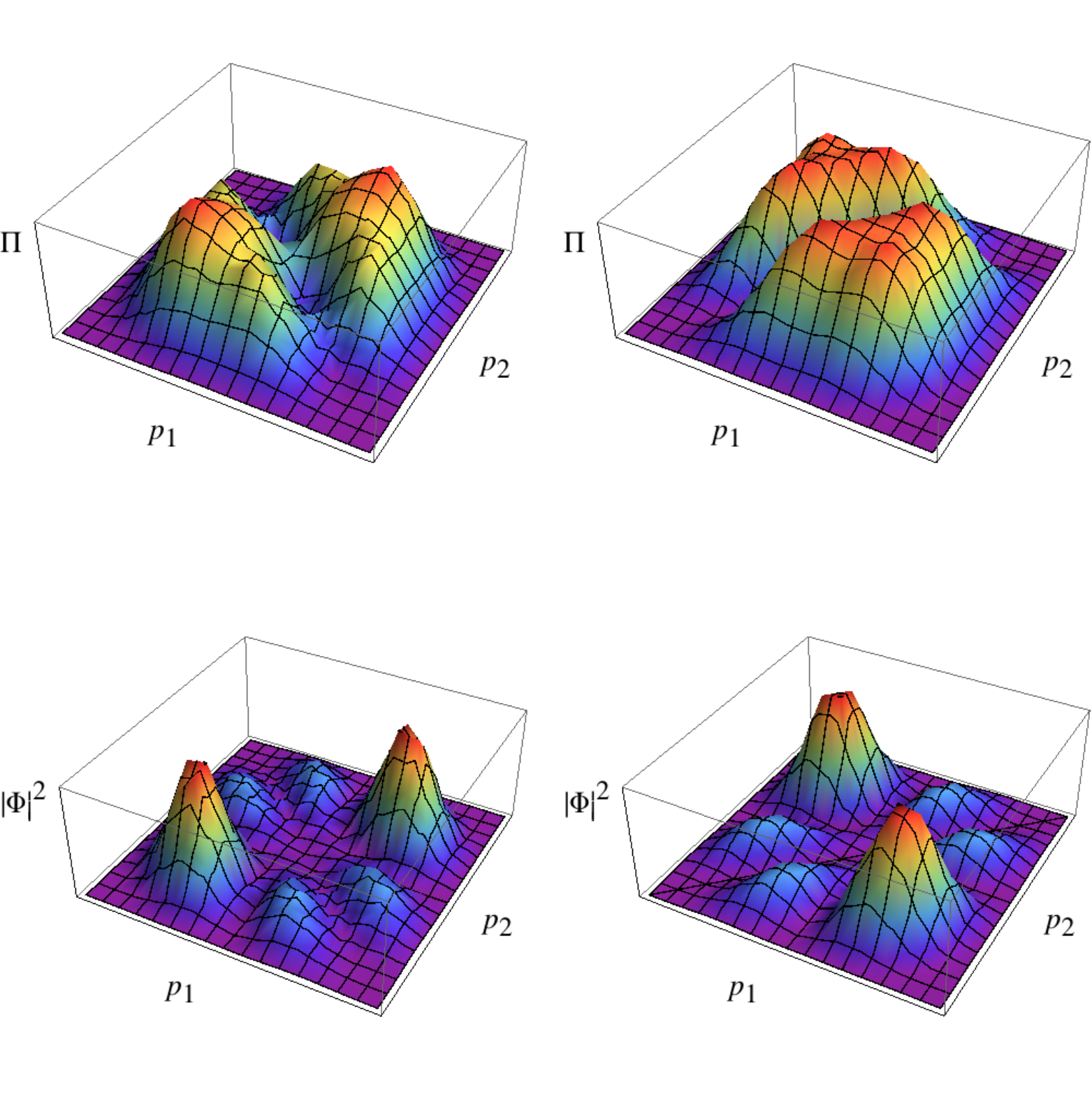}
\caption{\label{fig:distribucion momento HV} \small{Two-particle reduced densities in momentum space, $\Pi (p_1,p_2)$ (top), with $(n_1, n_2, n_3)$= $(1,2,3)$ and $L=1$ for symmetric (left) and antisymmetric (right) wave functions. The densities of the two-particle system, $|\Phi (p_1,p_2)|^2$ (bottom), with $(n_1, n_2)$=$(2,3)$ and $L=1$ for symmetric (left) and antisymmetric (right) wave functions, are also shown.}}
\end{figure}

We present in Figures \ref{fig:distribucion HV} and \ref{fig:distribucion momento HV} the marginal or two-particle reduced density distributions in position and in momentum space, $\Gamma (x_1,x_2)$ and $\Pi(p_1,p_2)$, for the symmetric and antisymmetric wave functions. The Fermi hole, where the 
density is zero when $x_1=x_2$ and $p_1=p_2$, is clearly present in the reduced distributions of the antisymmetric function. On the other hand, there is no such hole in the
densities of the symmetric function. 

This situation can be compared and contrasted to the densities of the two-particle system below where the presence of the Fermi hole is observed in the antisymmetric function.
There is also another symmetry hole in the symmetric function where the density is zero when $x_2=1-x_1$. This symmetry hole is due to the spatial symmetry of this system and
not a result of the imposed indistinguishability. This leads to equal values of two-particle entropies in the symmetric and antisymmetric functions when one quantum number is
even and the other is odd. As a consequence, their pair mutual information is the same \cite{laguna_jpa_2011}. Counterbalance holes which can also be present in symmetric wave functions have been reported and studied in atomic systems \cite{koga,koga1}.

This behavior is different from the three-particle system where the act of reducing (integrating over a variable) the density removes the presence of the symmetry hole. That is, $\Gamma (x,1-x)$ and $\Pi (p,1-p)$ are not equal to zero in the symmetric functions. We did not find any states for the three-particle system in which their two-particle entropies are equal for the symmetric and antisymmetric functions. Furthermore, we did not find any states where the three-particle entropies of the symmetric and antisymmetric functions are equal. 

One-particle (variable) reduced densities are not presented because they are the same for the symmetric and antisymmetric functions in the two-particle system \cite{laguna_jpa_2011}, and in these three-particle systems. These are
\begin{equation}
\rho (x)= \frac{1}{3} \sum\limits_{i=1}^3|\psi_{n_i}(x)|^2 \hspace{1cm}  \pi (p)= \frac{1}{3} \sum\limits_{i=1}^3 |\phi_{n_i}(p)|^2.
\label{onedens}
\end{equation}
Thus, symmetric and antisymmetric functions with the same quantum numbers have the same values of one-particle Shannon entropies. We do not present contour plots for the three-particle densities since it is difficult to visually ascertain differences between the functions.

\subsection{Mutual Information with two variables}

\begin{figure}
      \includegraphics[width=\columnwidth]{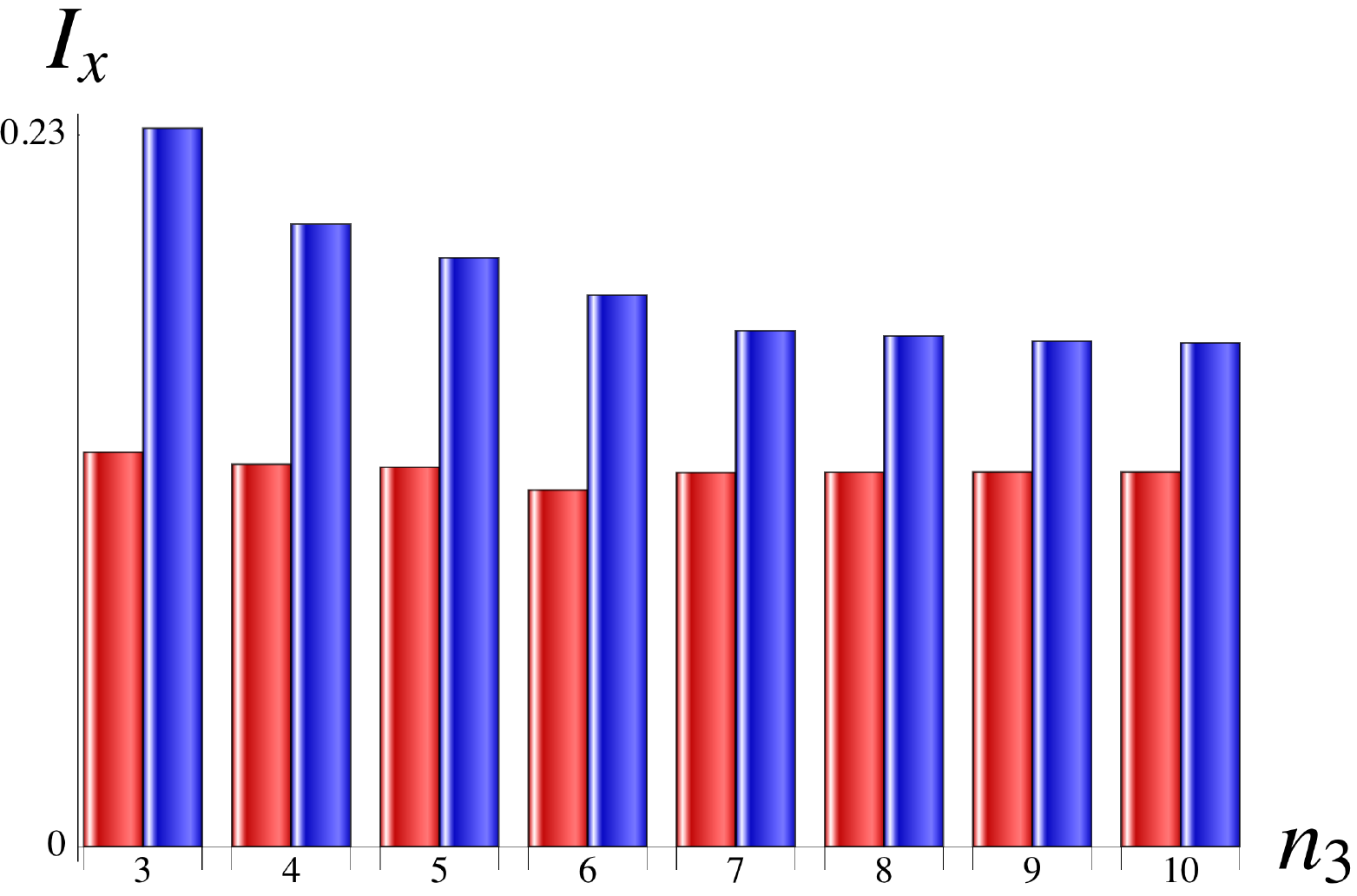}
       \caption{\label{fig:Iparesposicion} \small{Two-variable or pair mutual information, $I_x$, in position space for symmetric (red) and antisymmetric (blue) functions, varying $n_3$ with $n_1=1$, $n_2=2$ and $L=1$. Color online.}}
\end{figure}

\begin{figure}
      \includegraphics[width=\columnwidth]{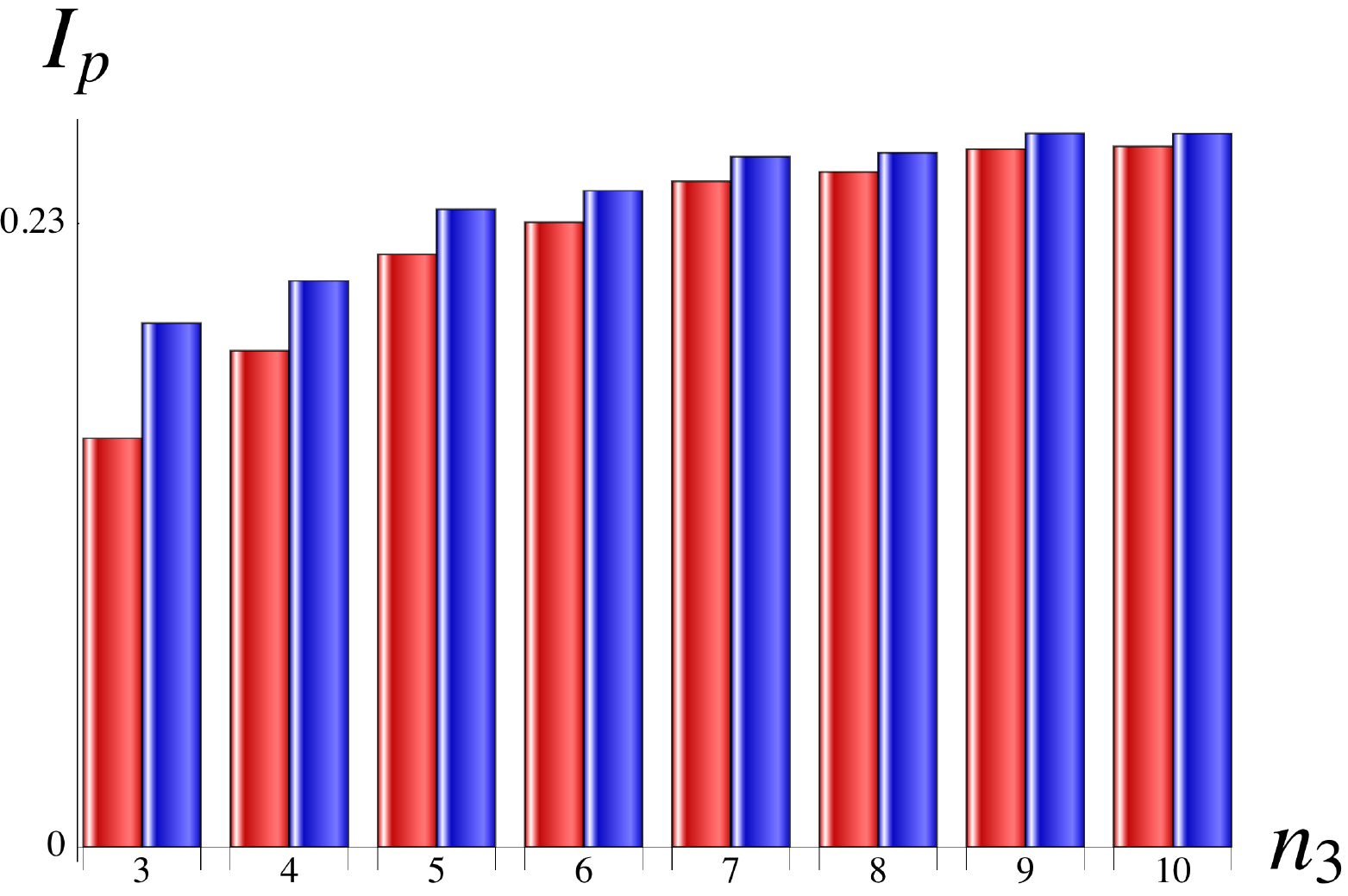}
       \caption{\label{fig:Iparesmoment} \small{Two-variable or pair mutual information, $I_p$, in momentum space for symmetric (red) and antisymmetric (blue) functions, varying $n_3$ with $n_1=1$, $n_2=2$ and $L=1$. Color online.}}
\end{figure}

The mutual information between pairs of variables (Eq. (\ref{mutinf})) is presented in Figures \ref{fig:Iparesposicion} and \ref{fig:Iparesmoment} for symmetric and antisymmetric functions in position and in momentum space. 
The results for the information measures are presented by choosing $n_1=1$, $n_2=2$ and varying the value of $n_3$.

It is clear that the value of mutual information and hence the statistical correlation, is larger for the antisymmetric functions as compared to the symmetric ones with same quantum numbers, as expected. This is due to the presence of the Fermi hole in the two-particle reduced distributions of the antisymmetric functions and the lack of a corresponding symmetry hole in the symmetric ones, as in the case of the two-particle systems when both quantum numbers have the same parity. These results also hold for all the studied cases with other choices of quantum numbers.

There are some differences between the position and momentum space behaviors. First, mutual information increases with $n_3$ in momentum space while it decreases (antisymmetric) or slowly varies (symmetric) in position space. Secondly, the difference between the values of the two types of functions decreases in both position and momentum space as $n_3$ increases. 

Comparing the previous two figures, one notes that both types of functions are more correlated in momentum space as compared to position space ($I_p > I_x$), with the exception of the antisymmetric function when $n_3=3$.

The differences between the behaviors of the position space information measures (decreasing tendency) with $n_3$ and the momentum space ones (increasing tendency) is due to the signs of the entropies. The $s_x$ entropies are negative valued but do not vary greatly with $n_3$. The pair entropies are negative but increase with $n_3$. The information
measures are positive valued. Since their first term ($s_x$) slowly varies with $n_3$, their behavior is determined by the absolute value of the pair entropy, which decreases.
On the other hand, all momentum space entropies are positive valued and increase with $n_3$. The increasing tendency observed in the information measures must be due primarily to increases in the $s_p$ first term.

\subsection{Mutual Information with three variables}

\begin{figure}
      \includegraphics[width=\columnwidth]{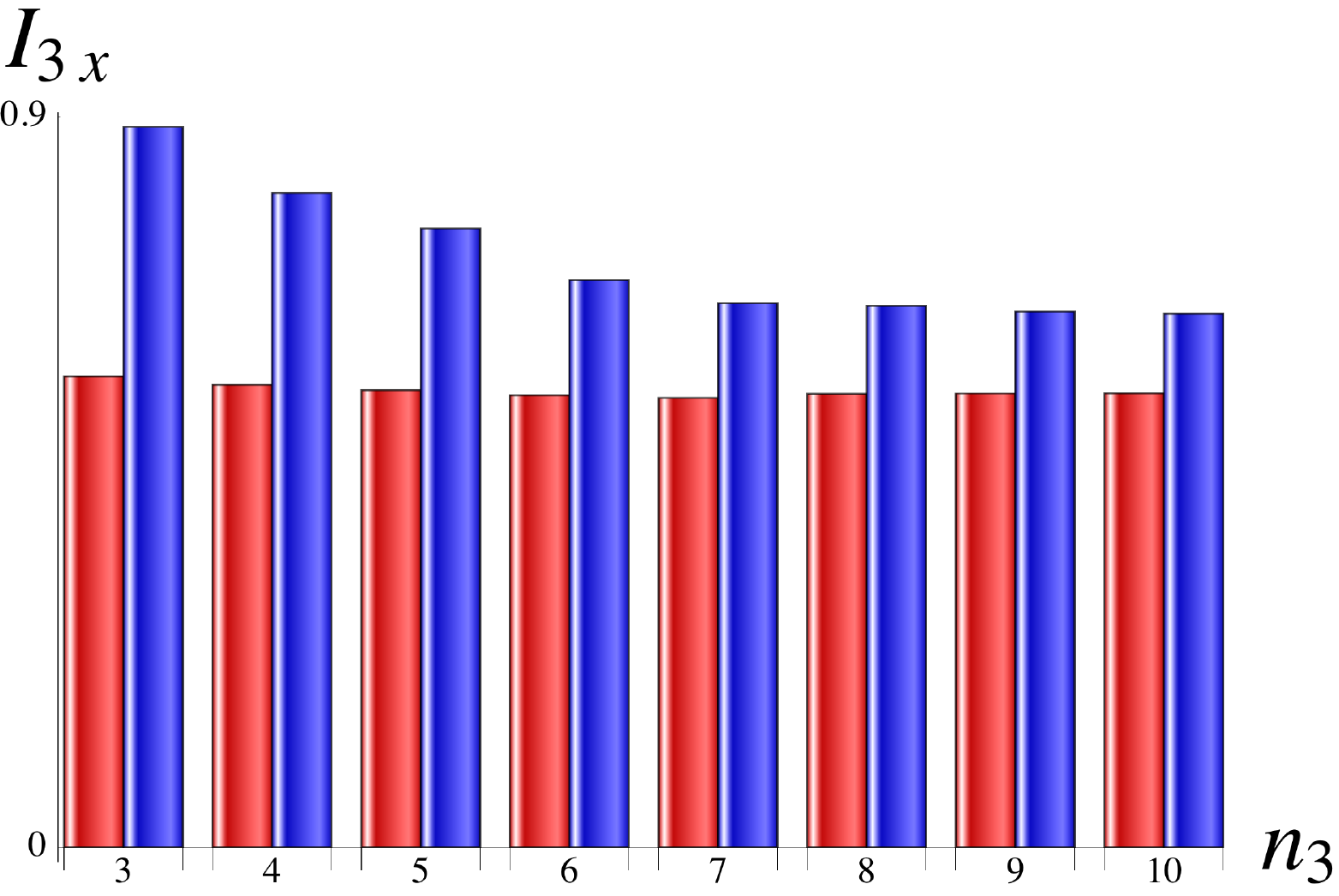}
       \caption{\label{fig:Ix} \small{Three-variable mutual information, $I_{3x}$, in position space for symmetric (red) and antisymmetric (blue) functions, varying $n_3$ with $n_1 = 1$, $n_2 = 2$ and $L=1$. Color online.}}
\end{figure}

\begin{figure}
      \includegraphics[width=\columnwidth]{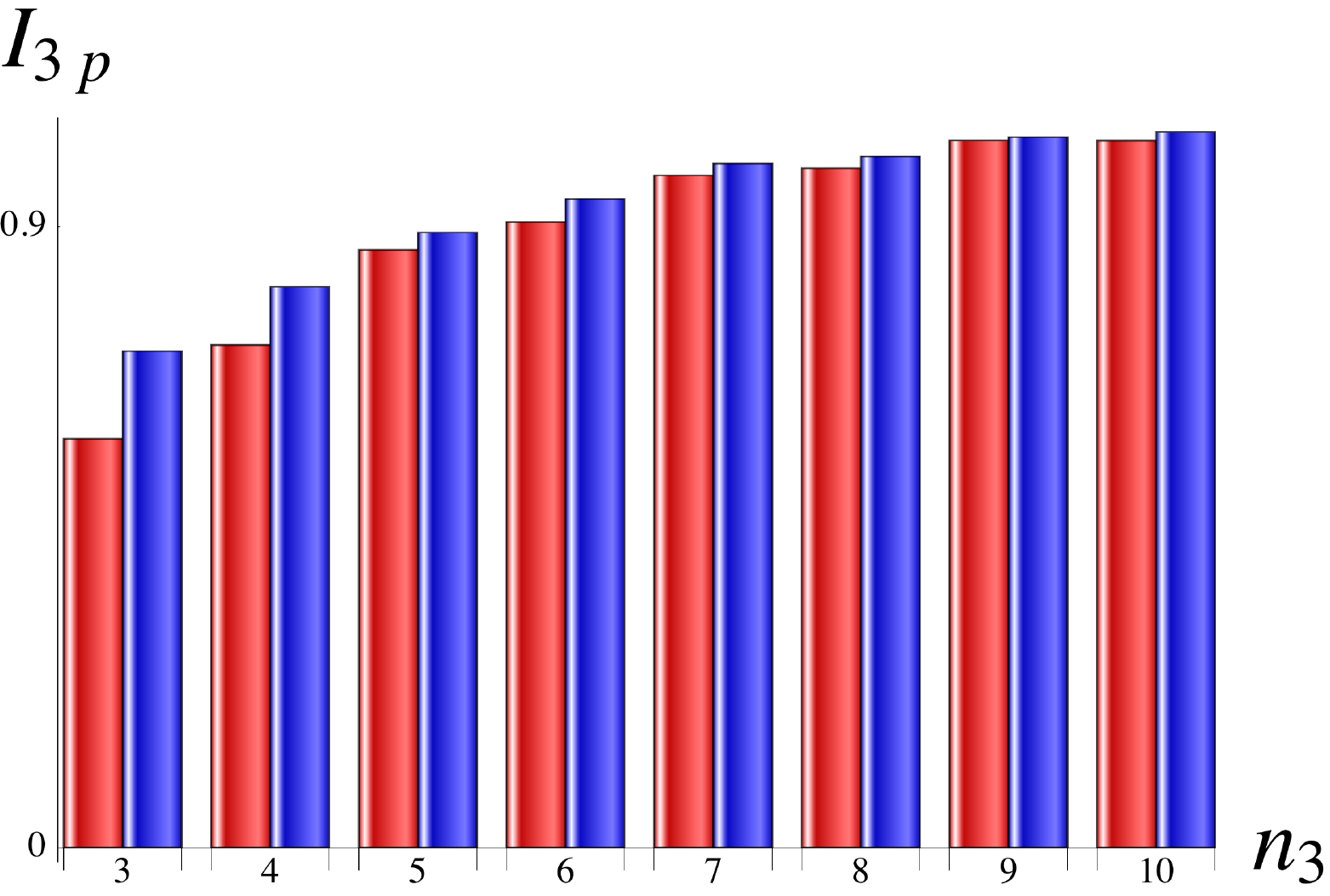}
      \caption{\label{fig:Ip} \small{Three-variable mutual information, $I_{3p}$, in momentum space for symmetric (red) and antisymmetric (blue) functions, varying $n_3$ with $n_1 = 1$, $n_2 = 2$ and $L=1$. Color online.}}
\end{figure}

Mutual information between the three variables from Eqs. (\ref{mi1x}) and (\ref{mi1p}) are presented in Figures \ref{fig:Ix} and \ref{fig:Ip} for position and momentum space. We do not present figures for the two other three-variable measures (Eqs. (\ref{mi2x}) - (\ref{mi3p})) for the sake of brevity. The general trends observed in Figures \ref{fig:Ix} and \ref{fig:Ip} are characteristic of the other measures.

The behavior in Figures \ref{fig:Ix} and \ref{fig:Ip} is very similar to that observed between the pairs in Figures \ref{fig:Iparesposicion} and \ref{fig:Iparesmoment}. That is, the antisymmetric functions are more correlated than the symmetric ones. This is consistent with the presence of the Fermi hole in the antisymmetric functions, and the lack of a symmetry hole in the symmetric ones. 

The difference between the two types of functions is smaller in momentum space as compared to position space. Correlation increases with $n_3$ in momentum space while it decreases or slowly varies in position space. 

Similar to the behavior observed for the pair correlations, $I_{3p} > I_{3x}$, for both types of functions with the exception of the $n_3=3$ antisymmetric state. 
\subsection{Higher-order mutual information}

\begin{figure}
      \includegraphics[width=\columnwidth]{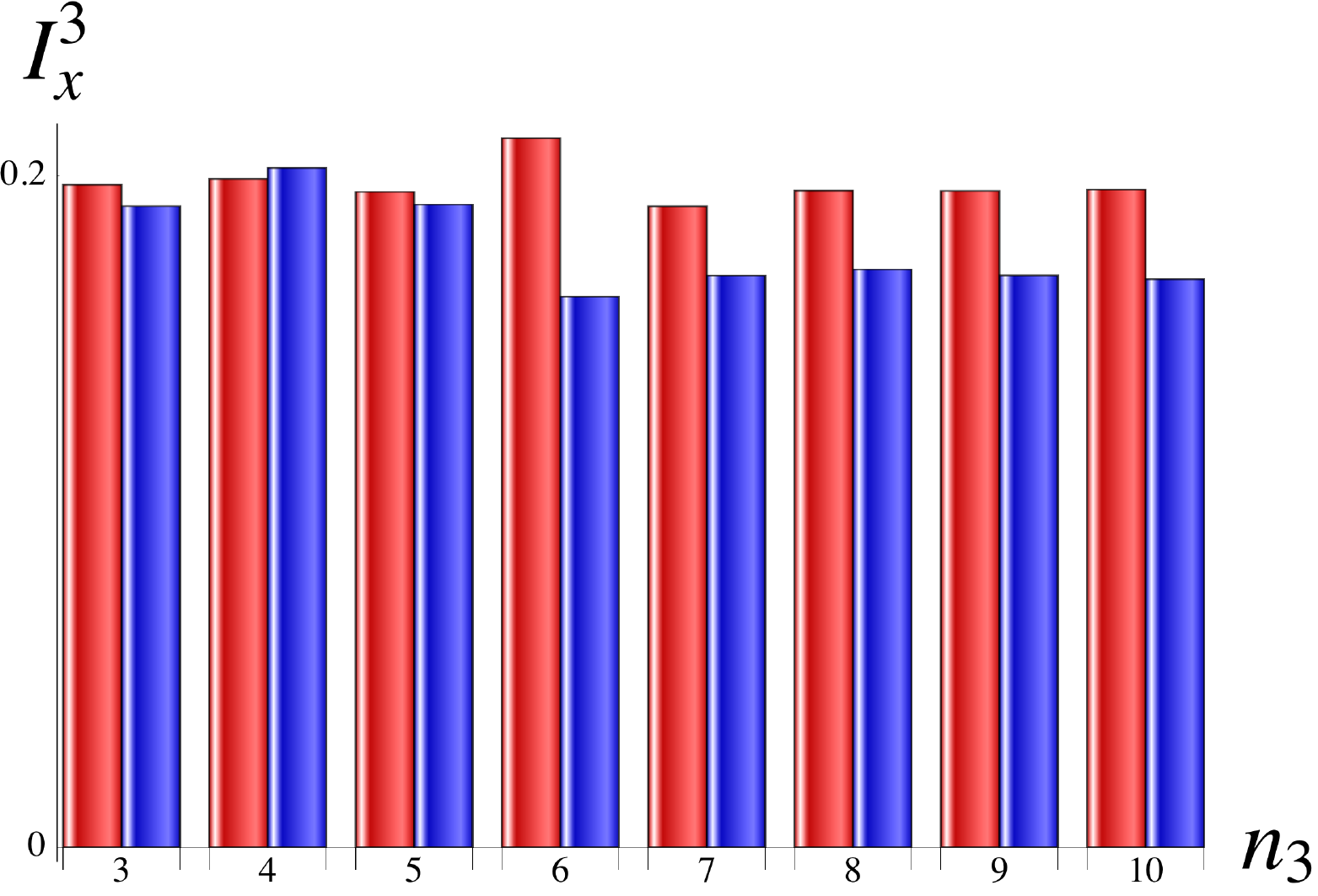}
       \caption{\label{fig:Ix1;x2;x3} \small{Higher-order mutual information, $I_{x}^3$, in position space for symmetric (red) and antisymmetric (blue) functions, varying $n_3$ with $n_1 = 1$, $n_2 = 2$ and $L=1$. Color online.}}
\end{figure}

\begin{figure}
      \includegraphics[width=\columnwidth]{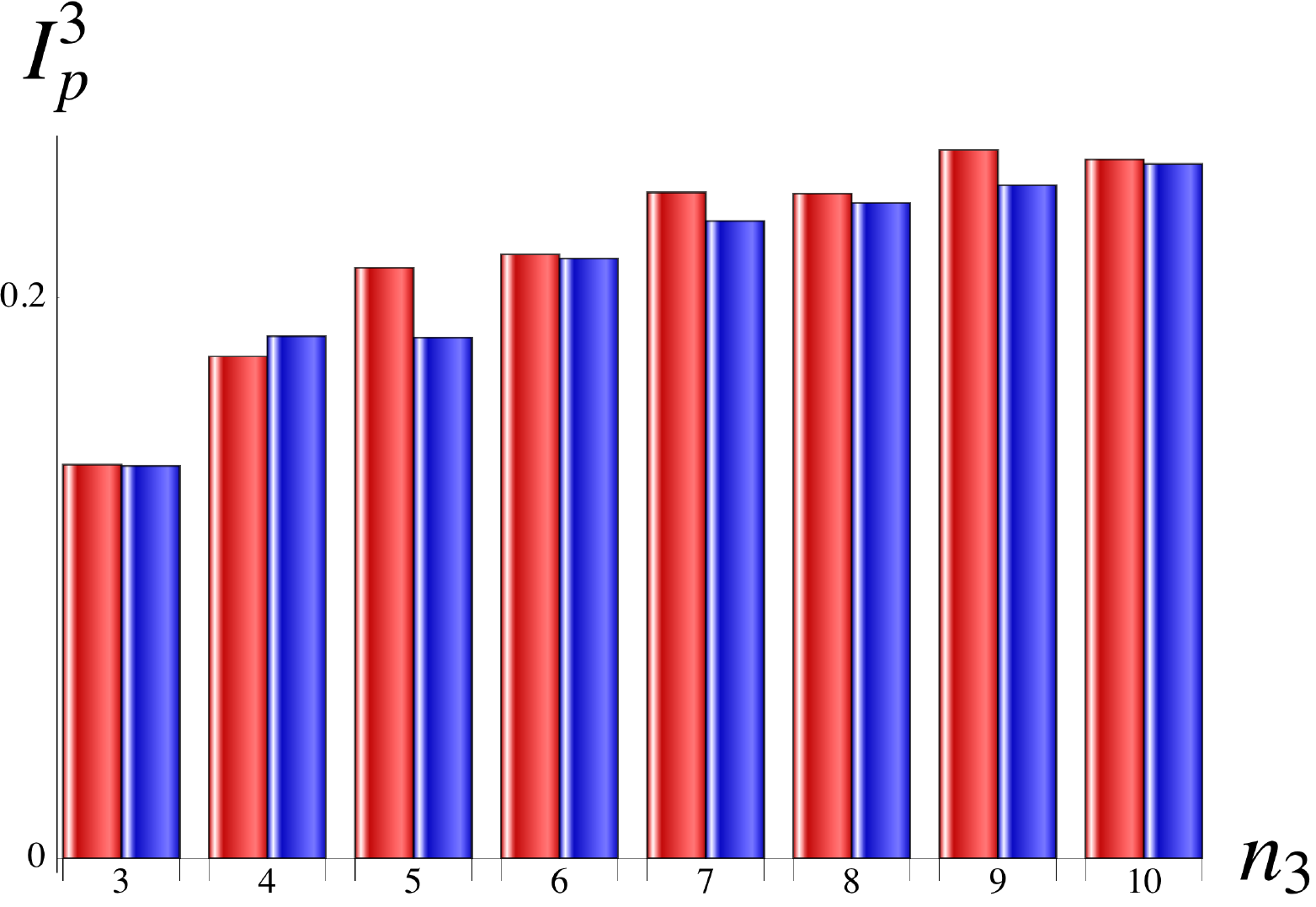}
      \caption{\label{fig:Ip1;p2;p3} \small{Higher-order mutual information, $I_{p}^3$, in momentum space for symmetric (red) and antisymmetric (blue) functions, varying $n_3$ with $n_1 = 1$, $n_2 = 2$ and $L=1$. Color online.}}
\end{figure}

Plots of the higher-order mutual information (Eqs. (\ref{mi4x}) and (\ref{mi4p})) in position and momentum space are presented in Figures \ref{fig:Ix1;x2;x3} and \ref{fig:Ip1;p2;p3}. 

These figures illustrate that it is now the symmetric functions which are more correlated than the antisymmetric ones in both spaces. This behavior is different to that observed between pairs of variables and the three-variable mutual information where the antisymmetric functions are more correlated. The only deviation from this trend occurs for the $n_3 = 4$ state in both position and momentum spaces. 

However, the behavior of higher-order mutual information is similar to the other measures in that the magnitude of the correlation increases with $n_3$ in momentum space and the differences between the two types of functions decrease with $n_3$. This is different from position space where the differences between the two types of functions is more marked as $n_3$ increases. 

Comparing the magnitudes of the correlation in both spaces, one observes that there is larger correlation in position space up to $n_3 = 4$ for the symmetric functions and $n_3=5$ for the antisymmetric functions. For larger $n_3$, there is more correlation in momentum space. 

Plots of the three-particle Shannon entropies are not presented for the sake of brevity. They show in all cases that the value of the Shannon entropies is smaller for the antisymmetric functions as compared to the corresponding symmetric ones with the same quantum numbers. The interpretation is that the three-particle densities of the antisymmetric functions  are more localized  than the symmetric ones. This behavior holds for position as well as momentum space. Furthermore, this behavior was also observed for the two-particle entropies. That is, the reduced two-particle densities from the antisymmetric functions are more localized than the symmetric ones (see next section). 

Note that although the behavior of all entropies is consistent with the presence of the Fermi hole in the antisymmetric functions, the combination of these entropies to form the higher-order mutual information, through Eqs. (\ref{mi4x}) and (\ref{mi4p}), yields that the symmetric functions are more correlated.
Furthermore, any differences in the higher-order mutual information are a result of differences between two and three-particle Shannon entropies since one-particle entropies are equal for both types of functions.

\subsection{Numerical data analysis}
The trends and observations in the sections above can be summarized and further understood by an analysis of the data.
\begin{table}[H]
\begin{tabular}{|c||c|c||c|c||c|c||c|c|}
\hline
		&	A&	S	&	A&	S	&	A&	S	&	A&	S	\\
$n_3$&	3	&	3		&	4		&	4		&	5		&	5		&	6		&	6		\\ \hline \hline
$s_x$ &  -0.1194 &  -0.1194 &  -0.1115 &  -0.1115 &  -0.1047 &  -0.1047
 &  -0.1024 &  -0.1024 \\
$s_{\Gamma}$ &  -0.4709 &  -0.3663 &  -0.4241 &  -0.3465 &  -0.3996 &  -0.3320
 &  -0.3831 &  -0.3201 \\
$s_{\Psi}$  &  -1.2455 &  -0.9382 &  -1.1403 &  -0.9042 &  -1.0763 &  -0.8772
 &  -1.0059 &  -0.8641 \\
$I_x$ &   0.2321 &   0.1275 &   0.2011 &   0.1235 &   0.1903 &   0.1226
 &   0.1782 &   0.1152 \\
$I_{3x}$ &   0.8872 &   0.5799 &   0.8058 &   0.5697 &   0.7623 &   0.5631
 &   0.6986 &   0.5569 \\
$I_{\rho,\Gamma}$ &   0.6552 &   0.4524 &   0.6047 &   0.4461 &   0.5720 &   0.4405
 &   0.5204 &   0.4416 \\
$I_{\Gamma,\Gamma}$ &   0.4231 &   0.3249 &   0.4035 &   0.3226 &   0.3818 &   0.3179
 &   0.3422 &   0.3264 \\
$I^3_x$ &   0.1910 &   0.1974 &   0.2024 &   0.1991 &   0.1915 &   0.1953
 &   0.1639 &   0.2112 \\

\hline
\end{tabular}
\caption{Entropies and mutual information measures for some $(n_{1}=1,n_{2}=2,n_3)$ states. The A columns correspond to values from the antisymmetric functions while the S columns are values from the symmetric functions.}
\label{tablex}
\end{table}

Table \ref{tablex} presents values of the entropies and information measures for some states in position space. First, note that all entropies are negative-valued. Second, the reduced pair entropies, $s_{\Gamma}$, and the three-particle entropy, $s_{\Psi}$, are smaller in the antisymmetric cases, consistent with the interpretation that the antisymmetric functions are more localized than the symmetric ones with same quantum numbers. Third, the reduced one-variable entropies are the same for the symmetric/antisymmetric pairs with the same quantum numbers.

The $I_x$ and $I_{3x}$ measures depend on differences between $s_x$ and $s_{\Psi}$. $s_x$ is the same for a symmetric/antiymmetric pair, thus the behavior of $I_x$ and $I_{3x}$ is determined by the magnitude of $s_{\Psi}$. This is largest (most negative) in the antisymmetric function hence the values of $I_x$ and $I_{3x}$ are larger than those of the symmetric functions. According to these measures, the antisymmetric functions are more correlated than the corresponding symmetric ones.

The remaining information measures depend on $s_{\Gamma}$. $I_{\rho,\Gamma}$ depends on a sum between $s_x$ and $s_{\Gamma}$, and the difference between this result and $s_{\Psi}$. Since $s_x$ is the same for a symmetric/antisymmetric pair, the behavior of $I_{\rho,\Gamma}$ is due to the difference between $s_{\Gamma}$ and $s_{\Psi}$. 

For $I_{\Gamma,\Gamma}$ and $I^3_x$, the situation is even more complex since the measures depend on differences between $s_{\Gamma}$ and $s_x$ (which changes the sign), and the resulting difference with $s_{\Psi}$. 
Of all the states studied, we found that $I_x$, $I_{3x}$ and $I_{\rho,\Gamma}$ are all larger in the antisymmetric functions as compared to the symmetric ones. The interpretation here from these measures is that the antisymmetric functions are more correlated. 

In $I_{\Gamma,\Gamma}$, there are a few states which presented values which are larger in the symmetric functions, however the antisymmetric values are larger than the corresponding symmetric ones in the large majority of the studied cases. 

The greater part of the considered states yield $I_{3_x}$ values which are larger for the symmetric functions. The interpretation here is that there are a number of states where the symmetric functions are more correlated than the antisymmetric ones when higher-order correlations are taken into account, which presents a different behavior from the pair measures.

\subsection{Three uncoupled oscillators}

We have also studied the system of three noninteracting or uncoupled harmonic oscillators (HO) where the respective densities are built from orbitals
\begin{equation}
\psi ^{HO}_n(x)=\frac{1}{\sqrt{2^nn!}}{\bigg (}\frac{\omega}{\pi}{\bigg )}^{1/4}e^{-\omega x^2/2}H_n {\bigg (}\sqrt{\omega}x{\bigg )}, \qquad -\infty \le x \le \infty, \qquad n=0,1,\cdots \\
\label{ho}
\end{equation}
in an analogous manner to those presented for the PIAB system. $H_n(x)$ is the $n^{th}-$order Hermite polynomial and $\omega$ is the strength of the potential. 

\begin{table}[H]
\begin{tabular}{|c||c|c||c|c||c|c||c|c|}
\hline
		&	A&	S	&	A&	S	&	A&	S	&	A&	S	\\
$n_3$&	2	&	2		&	3		&	3		&	4		&	4		&	5		&	5		\\ \hline \hline
$s_x$ &  1.5769 &  1.5769 &  1.6844 &  1.6844 &  1.7563 &  1.7563
 &  1.8039 &  1.8039 \\
$s_{\Gamma}$ &  2.9423 &  3.0352 &  3.1593 &  3.2576 &  3.3230 &  3.3819
 &  3.4293 &  3.4781 \\
$s_{\Psi}$  &  3.9337 &  4.1972 &  4.2040 &  4.5344 &  4.5282 &  4.6893
 &  4.7076 &  4.7956 \\
$I_x$ &   0.2115 &   0.1186 &   0.2095 &   0.1112 &   0.1897 &   0.1307
 &   0.1786 &   0.1298 \\
$I_{3x}$ &   0.7970 &   0.5335 &   0.8492 &   0.5189 &   0.7408 &   0.5796
 &   0.7042 &   0.6163 \\
$I_{\rho,\Gamma}$ &   0.5855 &   0.4149 &   0.6397 &   0.4076 &   0.5511 &   0.4489
 &   0.5256 &   0.4865 \\
$I_{\Gamma,\Gamma}$ &   0.3740 &   0.2963 &   0.4302 &   0.2964 &   0.3615 &   0.3181
 &   0.3470 &   0.3567 \\
$I^3_x$ &   0.1624 &   0.1778 &   0.2208 &   0.1852 &   0.1718 &   0.1874
 &   0.1684 &   0.2268 \\
\hline
\end{tabular}
\caption{Entropies and mutual information measures for some $(n_{1}=0,n_{2}=1,n_3)$ states of the three uncoupled oscillators. The A columns correspond to values from the antisymmetric functions while the S columns are values from the symmetric functions.}
\label{tablex1}
\end{table}

Table \ref{tablex1} reports values of the entropies and information measures for some states. The value of $\omega$ was chosen to be unity since the values of the entropies and information measures are equal in position and in momentum space. Larger values of $\omega$ localize the densities in position space which results in smaller entropic values. 
On the other hand, the densities delocalize in momentum space, with larger entropic values for larger $\omega$.
However, the values of the mutual information measures are independent of the value of $\omega$. 

Note that all entropies are now positive-valued, in contrast to the values presented for the PIAB system.
Nevertheless, the trends are consistent with those observed in the PIAB system. That is, $s_{\Gamma}$ and $s_{\Psi}$ are smaller for the antisymmetric functions, while all entropies increase with $n_3$. Most importantly,  $I_{3_x}$, the higher-order measure,  is larger for the symmetric functions in the majority of cases, in contrast to the other measures which are larger for the antisymmetric functions. The behavior of the information measures with $n_3$ is different to that observed in PIAB. The $I_x$ measure for
the antisymmetric functions is the only one that shows a decrease with $n_3$. All other measures exhibit both increasing and decreasing tendencies.

\section{Superposition of states}

In this section, we analyze the behavior of higher-order mutual information as a correlation measure. In order to do this, one requires a model where correlation can be adjusted in a systematic manner to gauge its effect on the measure. In the absence of potentials, one can conceive of introducing correlation via the use of a superposition of states.  The corresponding function in position space is
\begin{equation}
\Psi_S (x_1,x_2,x_3) = c_1 \; \Psi_{n_1,n_2,n_3}(x_1,x_2,x_3)  + c_2 \; \Psi_{n_4,n_5,n_6}(x_1,x_2,x_3).
\label{superfunx}
\end{equation}
The three-particle density, $|\Psi_S (x_1,x_2,x_3)|^2$, can then be used to calculate the three-particle Shannon entropy, and reduced by integration, to obtain reduced or marginal densities from which the reduced Shannon entropies can be calculated. These expressions have been given above. 

In such functions, a further source of correlation beyond the indistinguishability, is the quantum interference or cross terms between states, that are present in the densities.
This type of correlation is distinct from the one due to indistinguishability. The coefficients, $|c_1|^2 + |c_2|^2=1$, can be varied to change the correlation in the system. We restricted ourselves to real-valued coefficients. 

Figure \ref{fig:Ix,Ix3} presents values of the third-order and pair  mutual informations in position space as $c_1^2$ is varied from zero to one. We chose to study the particular case of
the superposition of 
\begin{equation}
\Psi_S (x_1,x_2,x_3)= c_1 \; \Psi_{1,2,3}(x_1,x_2,x_3) + c_2 \; \Psi_{4,5,6}(x_1,x_2,x_3).
\label{supernums}
\end{equation}

\begin{figure}
\includegraphics[width=\columnwidth]{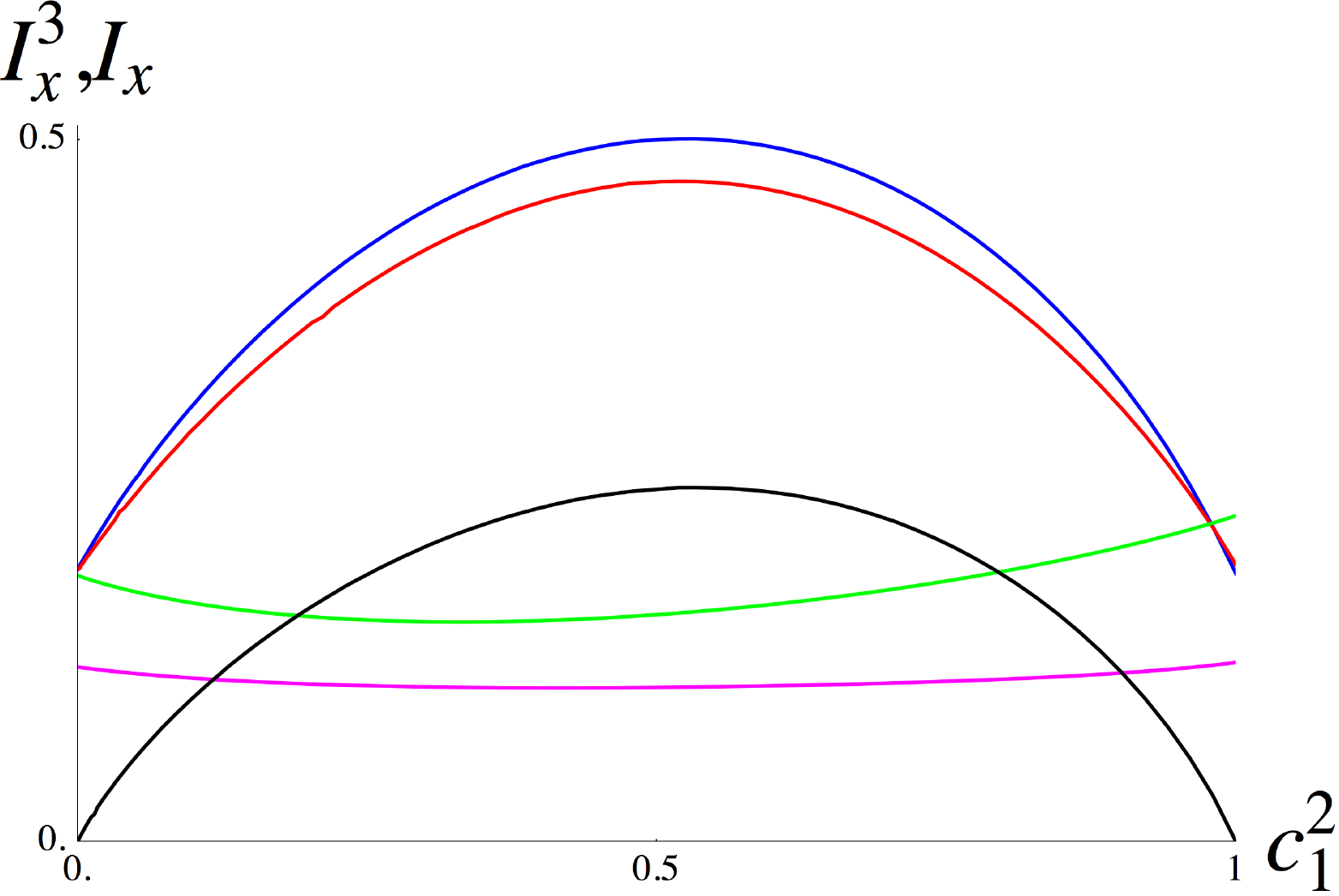}
\caption{\label{fig:Ix,Ix3} \small{Higher-order, $I_x^3$ (symmetric, red), (antisymmetric, blue), and pair, $I_x$, mutual informations in position space, from the superposition function, $\Psi_S(x_1,x_2,x_3)$, for symmetric (pink) and antisymmetric (green) functions, varying $c_1^2$. $I_x^3$ for the superposition of distinguishable functions, $\Psi_S^D(x_1,x_2,x_3)$, (black) is also shown. Color online.}}
\end{figure}
What is clear is that $I_x^3$ is sensitive to the correlation and attain its maximum at $c_1^2=0.5$. This corresponds to a balanced superposition state where $c_1=c_2=\frac{1}{\sqrt{2}}$. 

Also interesting in the consideration of these superpositions is that the antisymmetric functions are now more correlated than the symmetric ones. This is opposite to the behavior previously observed for one state. 

Plots of the pair mutual information are also included in the figure. The results illustrate that this measure is
insensitive to the correlation due to the quantum interferences and actually decrease with $c_1^2$ when one would expect increases due to added correlation. 

$I_x^3$ values for superpositions of distinguishable functions, $ \Psi_S^D (x_1,x_2,x_3)$, (Hartree-like product functions with no symmetrization)

\begin{equation}
\Psi_S^D (x_1,x_2,x_3)= c_1 \; \psi_1(x_1)\psi_2(x_2)\psi_3(x_3) + c_2 \; \psi_4(x_1)\psi_5(x_2)\psi_6(x_3),
\label{superdis}
\end{equation}
are also presented. One can observe an order in the correlation with the distinguishable functions being the least correlated followed by the symmetric ones and then the antisymmetric ones. This relative ordering holds throughout the range of $c_1^2$.
Note that $I_x^3$ is zero when there is no superposition ($c_1^2$ equal to zero or one) since there is no correlation from indistinguishability in these distinguishable functions.

The higher-order mutual information was further tested by examining the correlation generated from states where the quantum interferences are not included (no $c_1c_2$ cross terms in the densities). These results are presented in Figure \ref{fig:Ix1,x2,x3M}. 

\begin{figure}
      \includegraphics[width=\columnwidth]{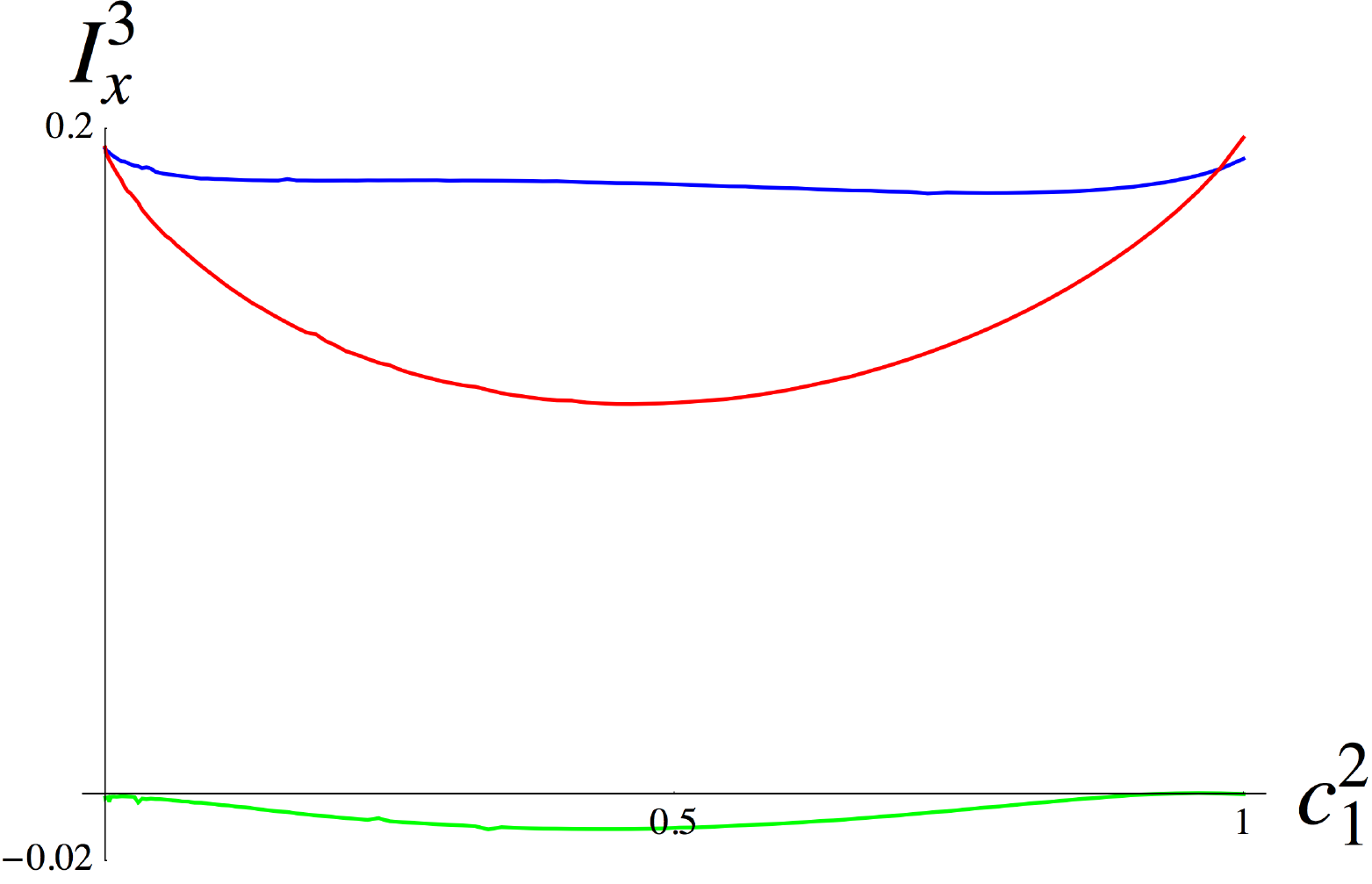}
      \caption{\label{fig:Ix1,x2,x3M} \small{Higher-order mutual information, $I_x^3$, in position space, from the superposition function, $\Psi_S(x_1,x_2,x_3)$, without quantum interferences, for symmetric (red) and antisymmetric (blue) functions, varying $c_1^2$. $I_x^3$ for superpositions of the distinguishable functions, $\Psi_S^D(x_1,x_2,x_3)$, (green) without quantum interferences, is also shown. Color online.}}
\end{figure} 
It is apparent that the correlation in the antisymmetric functions is now not as sensitive to the value of $c_1^2$. This can be contrasted against the behavior in the previous figure and shows that the third-order measure is sensitive to the correlation from the quantum interferences. There is a minimum in the symmetric functions instead of the maximum in the previous figure. However, the antisymmetric functions are more correlated than the symmetric ones ( 0 $<$ $c_1^2$ $<$ 1) as in the previous figure where quantum interferences are included. When $c_1^2$  is equal to zero or one, the symmetric functions are more correlated.

Results from the distinguishable functions show negative values for the third-order measure  and can be contrasted to the positive values which occur when indistinguishability is imposed, or in the previous figure when quantum interferences are included. This suggests that the positivity or negativity of the third-order mutual information depends on physical constraints imposed on the system. Furthermore, the minimum where $I_x^3$ is maximally negative, occurs when  $c_1=c_2=\frac{1}{\sqrt{2}}.$

The interpretation in momentum space is less clear and the plots are not presented. The reason is that the quantum interferences in momentum space do not contribute to a large extent in the overall correlation.
It is now the mutual information between pairs which exhibit maximal correlation at the balanced superposition state. On the other hand, the third-order measures exhibit minima instead of the maxima seen in the previous figures. 
Similar to position space, the distinguishable functions without quantum interferences yield negative values.

\section{Conclusions}

Higher-order mutual information is used to quantify and study the higher-order statistical correlation present in a system of three spinless non-interacting quantum particles in a unidimensional box, in position and in momentum space. The statistical correlation in these systems arises from the indistinguishability requirements on the wave function. The results show that the magnitude of the higher-order mutual information is larger for the symmetric wave functions, in comparison to the antisymmetric ones with the same quantum numbers, in the majority of the states studied. This is different from other three-variable measures and the two-variable (pair) mutual information where the antisymmetric functions are in general more correlated. It is also distinct from similar two-particle systems where the antisymmetric functions are more (or equally) correlated than the corresponding symmetric ones. These results are also consistent with those observed in a system of three uncoupled oscillators. Indeed, more particles lead to a different behavior \cite{anderson}.
The utility of higher-order mutual information as a correlation measure is explored by considering a superposition of states where there is correlation due to quantum interferences. We observe that the magnitude of the higher-order mutual information is sensitive to the superposition coefficients with the maximum value attained for the balanced superposition state. This provides evidence that the magnitude can be used as a measure of correlation strength. Furthermore, the inclusion of superposition induces the antisymmetric functions to now be more correlated than the symmetric ones, with both being more correlated than distinguishable functions. The pair mutual information is less sensitive to the correlations arising from the quantum interferences due to the use of reduced densities in its definition. Higher-order mutual information arising from superpositions of distinguishable functions without quantum interferences is negative-valued. This is in contrast to the positive values obtained when the quantum interferences are included, and to the results from superpositions of symmetric and antisymmetric functions. These results suggest a dependence of the sign of higher-order mutual information on physical constraints in these systems.

\section*{Acknowledgements}

V.S.Y. thanks CONACyT for a fellowship. H.G.L. thanks DGAPA-UNAM for a postdoctoral fellowship. The authors acknowledge CONACyT for support to the
Red de Fisicoqu\'imica Te\'orica (RedFQT) through project 0250976.

\end{document}